\documentclass[11pt,a4paper]{article}
\usepackage[margin=1in]{geometry}
\usepackage{pgfplots}
\usepackage{hyperref}
\usepackage{tikz}

\usepackage{amsmath}
\usepackage{amsfonts}
\usepackage{amssymb}
\usepackage[USenglish]{babel}
\usepackage{graphicx}
\usepackage{booktabs}
\usepackage{threeparttable}
\usepackage{array}
\usepackage{dcolumn}
\newcolumntype{d}{D{.}{.}{-1}}
\usepackage[utf8]{inputenc}

\providecommand{\nolinenumbers}{}

\newcolumntype{+}{!{\vrule width 2pt}}

\newlength\savedwidth

\newcommand\thickhline{\noalign{\global\savedwidth\arrayrulewidth\global\arrayrulewidth 2pt}%
\hline
\noalign{\global\arrayrulewidth\savedwidth}}

\providecommand\plos[1]{#1}

\begin{document}

\title{Business disruptions from social distancing}
\author{Miklós Koren and Rita Pető\thanks{Koren: Central European University, KRTK KTI and CEPR. 1051 Budapest, Nádor u. 9., Hungary. E-mail: korenm@ceu.edu. Pető: KRTK KTI. E-mail: rita.peto@krtk.mta.hu.}}

\date{March 2020}
\maketitle

\renewcommand{\baselinestretch}{1.25} \normalsize
\begin{abstract}
Social distancing interventions can be effective against epidemics but are potentially detrimental for the economy.
Businesses that rely heavily on face-to-face communication or close physical proximity when producing a product or providing a service are particularly vulnerable. 
There is, however, no systematic evidence about the role of human interactions across different lines of business and about which will be the most limited by social distancing.
Here we provide theory-based measures of the reliance of U.S. businesses on human interaction, detailed by industry and geographic location.
We find that 49 million workers work in occupations that rely heavily on face-to-face communication or require close physical proximity to other workers. Our model suggests that when businesses are forced to reduce worker contacts by half, they need a 12 percent wage subsidy to compensate for the disruption in communication. Retail, hotels and restaurants, arts and entertainment and schools are the most affected sectors.
Our results can help target fiscal assistance to businesses that are most disrupted by social distancing.
\end{abstract}

\section*{Introduction}
Social distancing measures are effective non-pharmaceutical interventions against the rapid spread of epidemics \cite{Bootsma2007-ww,Markel2007-ad,Hatchett2007-gc,Wilder-Smith2020-jj}. Many countries have implemented or are considering measures such as school closures, prohibition of large gatherings and restrictions on non-essential stores and transportation to slow down the spread of the 2019--20 coronavirus pandemic \cite{Anderson2020-qk,Cohen2020-jw,Thompson2020-lc,noauthor_2020-xi}. What are the economic effects of such social distancing interventions? Which businesses are most affected by the restrictions?

Past research has analyzed the efficacy of social distancing interventions on reducing the spread of epidemics using the 1918 Spanish Flu in the U.S. \cite{Hatchett2007-gc,Markel2007-ad,Bootsma2007-ww} and seasonal viral infections in France \cite{Adda2016-mn}. Our knowledge of economic impacts, however, is limited \cite{Wren-Lewis2020-vc}. For this question, past data may be less relevant, as the importance of face-to-face communication has increased steadily in the last 100 years through urbanization \cite{Henderson2010-mv,Henderson2002-ji} and specialization increased in business services as well \cite{Herrendorf2014-jx,Duarte2019-kc}. 

The starting point of this paper is the observation that many sectors rely heavily on face-to-face communication in the production process \cite{Charlot2004-zr,Tian2019-wq}. We build a model of communication to understand how limiting face-to-face interaction increases production costs. Without social distancing, workers specialize in a narrow range of tasks and interact with other workers completing other tasks. This division of labor reduces production costs but requires frequent contact between workers. In the model, the number of contacts per worker is the most frequent in high-population-density areas in businesses where the division of labor is important. When face-to-face interaction is limited, these are exactly the businesses that suffer the most.

To measure business disruptions from social distancing, we turn to recent data on the task descriptions of each occupation \cite{National_Center_for_ONET_Development2020-wj} and the precise geographic location of non-farm businesses in the U.S. \cite{CBP}. We construct three groups of occupations and study their distribution in space. First, some occupations require face-to-face communication several times a week with other workers. Examples of these \emph{teamwork-intensive} occupations include maintenance, personal care related occupations and health care professionals. Other occupations require frequent contact with customers. Counselors, social workers and salespersons are examples of such \emph{customer-facing} occupations. The third group of workers may need to be in physical proximity of one another even if they do not communicate, for example, to operate machinery or access key resources. Examples of such occupations requiring \emph{physical presence} are drivers and machine operators, especially in mining and water transport, where crammed working environments are common.

The spatial distribution of employment matters because face-to-face interactions are more frequent in dense cities \cite{Charlot2004-zr,Tian2019-wq}. So any social distancing intervention imposes disproportionately more limitation on urban sectors. We use our theory-motivated measures to estimate which sectors and which locations will be particularly hurt by social distancing.

\section*{A model of communication}
When workers communicate with others, they can divide labor more effectively. Production involves sequentially completing tasks indexed by $z\in[0,1]$. A single worker can do a range of tasks, but there is a benefit to specialization and division of labor \cite{Smith1778-qq,Becker1992-ac}. The labor cost of a worker completing $Z<1$ measure of tasks is $Z^{1+\gamma}/\gamma$, where $\gamma>0$ captures the benefits to the division of labor. As we show below, the higher the $\gamma$, the more specialized each worker will be in a narrower set of tasks. Without loss of generality, we normalize the wage rate of workers to one so that all costs are expressed relative to worker wages.

Once the range of tasks $Z$ is completed, the worker passes the unfinished product on to another worker. This has a cost of $\tau$, which can capture the cost of communicating and interacting across workers. The determinants of communication cost will be parametrized later. After all the tasks are completed, another step of communication with cost $\tau$ is needed to deliver the product to the customer. This cost leads to the Marshallian externality that firms want to be close to their customers and customers want to be close to their suppliers \cite{Marshall1920-ps,Krugman1991-gr}.

The firm will optimally decide how to share tasks between workers. The key trade-off is economizing on the cost of communication while exploiting the division of labor \cite{Becker1992-ac}. Let $n$ denote the number of workers involved in the production process. Because workers are symmetric, each works on $Z=1/n$ range of tasks before passing the work to the next worker. Production involves $n-1$ ``contacts'' (instances of communication) and there is an additional contact with the customer. 

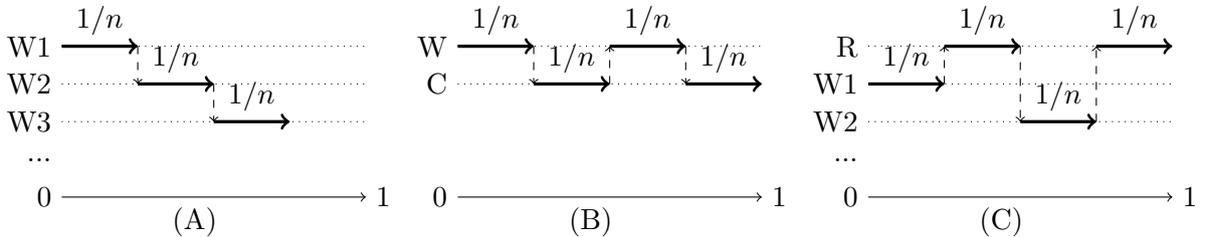
\begin{figure}[!h]
\begin{tikzpicture}
\draw [dotted] (0,1.5) node [left] {W1} --(4,1.5);
\draw [dotted] (0,1) node [left] {W2}--(4,1);
\draw [dotted] (0,0.5) node [left] {W3}--(4,0.5);
\node at (0,0) [left] {...};

\draw [very thick, ->] (0,1.5)--(0.5,1.5) node [above] {$1/n$} --(1,1.5);
\draw [very thick, ->] (1,1)--(1.5,1) node [above] {$1/n$}--(2,1);
\draw [very thick, ->] (2,0.5)--(2.5,0.5) node [above] {$1/n$}--(3,0.5);

\draw [dashed, ->] (1,1.5)--(1,1);
\draw [dashed, ->] (2,1)--(2,0.5);

\draw [->] (0,-0.5) node [left] {$0$} -- (1.75,-0.5) node [below] {(A)} --(4,-0.5) node [right] {$1$};
\end{tikzpicture}
\begin{tikzpicture}
\draw [dotted] (0,1.5) node [left] {W} --(4,1.5);
\draw [dotted] (0,1) node [left] {C}--(4,1);

\draw [very thick, ->] (0,1.5)--(0.5,1.5) node [above] {$1/n$} --(1,1.5);
\draw [very thick, ->] (1,1)--(1.5,1) node [above] {$1/n$}--(2,1);
\draw [very thick, ->] (2,1.5)--(2.5,1.5) node [above] {$1/n$}--(3,1.5);
\draw [very thick, ->] (3,1)--(3.5,1) node [above] {$1/n$}--(4,1);

\draw [dashed, ->] (1,1.5)--(1,1);
\draw [dashed, ->] (2,1)--(2,1.5);
\draw [dashed, ->] (3,1.5)--(3,1);

\draw [->] (0,-0.5) node [left] {$0$} -- (1.75,-0.5) node [below] {(B)} --(4,-0.5) node [right] {$1$};
\end{tikzpicture}
\begin{tikzpicture}
\draw [dotted] (0,1.5) node [left] {R} --(4,1.5);
\draw [dotted] (0,1) node [left] {W1}--(4,1);
\draw [dotted] (0,0.5) node [left] {W2}--(4,0.5);
\node at (0,0) [left] {...};

\draw [very thick, ->] (0,1)--(0.5,1) node [above] {$1/n$} --(1,1);
\draw [very thick, ->] (1,1.5)--(1.5,1.5) node [above] {$1/n$}--(2,1.5);
\draw [very thick, ->] (2,0.5)--(2.5,0.5) node [above] {$1/n$}--(3,0.5);
\draw [very thick, ->] (3,1.5)--(3.5,1.5) node [above] {$1/n$}--(4,1.5);

\draw [dashed, ->] (1,1)--(1,1.5);
\draw [dashed, ->] (2,1.5)--(2,0.5);
\draw [dashed, ->] (3,0.5)--(3,1.5);

\draw [->] (0,-0.5) node [left] {$0$} -- (1.75,-0.5) node [below] {(C)} --(4,-0.5) node [right] {$1$};
\end{tikzpicture}

\caption{{\bf Patterns of interaction in the workplace.}
Horizontal movement represents production, vertical movement represents interaction. (A) Each worker W works on a range $1/n$ of tasks, passing work $n-1$ times. (B) Worker W and customer C engage in frequent interactions. (C) Each worker W needs physical access to a key resource R.}
\label{fig1}
\end{figure}

Fig \ref{fig1} illustrates the division of labor between workers. Horizontal movement represents production along a range of tasks ($Z=1/n$), vertical movement represents interaction ($\tau$). We note three potential interpretations of our model. First, when workers work in teams, they can efficiently divide labor among themselves (panel A). The benefit of a larger team is better specialization. Second, communication may involve the customer (panel B). The benefit of more frequent interaction with the customer is a product or service that is better suited to their needs. Third, workers may need access to a key physical resource, such as a machine, vehicle or an oil well (panel C). In this case, even if they do not communicate, they may be subject to social distancing measures. The key assumption behind all three interpretations is that frequent interaction increases productivity, whether happening between workers, between workers and customers, or between workers and machines.

The firm's cost minimization problem can then be written as a function of the number of contacts alone,
\begin{equation}\label{TC}
	c(\tau) = \min_n n\tau + \frac 1\gamma n^{-\gamma},
\end{equation}
where total communication costs are $n\tau$ and production costs are $n Z^{1+\gamma}/\gamma$ with $Z=1/n$.

Given the strict convexity of this cost function, and ignoring integer problems, the first-order condition is necessary and sufficient for the optimum,
\begin{equation}\label{nstar}
	n^*(\tau) = \tau^{-1/(1+\gamma)}.
\end{equation}
The number of worker contacts is decreasing in the cost of communication, expressed relative to worker wage. When the division of labor is important, $\gamma$ is high, and the number of contacts does not depend very strongly on communication costs.

The total cost of producing one good can be calculated by substituting in \eqref{nstar} into \eqref{TC},
\begin{equation}\label{cost}
	c(\tau) = \tau^{\chi}/\chi,
\end{equation}
where $\chi=\gamma/(1+\gamma)\in(0,1)$ measures the importance of division of labor. This unit cost function is the same as if workers and communication were substitutable in the production function in a Cobb-Douglas fashion. Indeed, $\chi$ captures the share of costs associated with communication and can be calibrated accordingly.

\subsection*{The cost of communication}
Businesses may differ in $\chi$, how important communication is in their production process. Another source of heterogeneity is the cost of communication between workers, which we specify further as follows.

When workers meet face to face, communication costs depend inversely on the population density in the neighborhood of the firm, $\tau=d^{-\varepsilon}$ with $\varepsilon>0$. This captures the Marshallian externality of knowledge spillovers \cite{Marshall1920-ps}, which happen more easily in densely populated areas \cite{Charlot2004-zr,Rossi-Hansberg2007-tm,Ioannides2008-bs,Tian2019-wq}. Contacts will be more frequent in dense areas,
\begin{equation}\label{communication}
	n^*(d) = d^{\varepsilon(1-\chi)}.
\end{equation}
The unit cost of production is
\begin{equation}\label{cost2}
c(d) = d^{-\varepsilon\chi}/\chi.
\end{equation}
Firms in dense areas face lower unit costs \cite{Ciccone1996-gu}, but this agglomeration benefit may be offset by higher wages and land rents in places with high population density \cite{Eberts1982-zf,Madden1985-pq,Combes2019-ud}. In spatial equilibrium (not modeled here), firms with high communication needs will choose to locate in high-density areas \cite{Tian2019-wq}.

\subsection*{Social distancing measures}
We study the effect of a social distancing intervention that puts an upper limit $N$ on the number of face-to-face contacts. Firms can mitigate the disruption from this measure by moving communication online, but this is more costly per contact than face-to-face communication.

The optimal number of contacts without social distancing is given by Eq \eqref{communication}. Firms with $n^*>N$ are limited by social distancing. Without moving communication online, their unit cost will increase to $c' = N d^{-\varepsilon} + N^{-\gamma}/\gamma$, which is greater than the optimal cost,
\begin{equation}\label{socdist}
\frac{c'(d)}{c(d)} 
	= \chi \frac{N}{n^*(d)} + (1-\chi) \left(\frac{N}{n^*(d)}\right)^{-\gamma}>1.
\end{equation}
The first term of the weighted average is less than one, representing a reduction in communication costs once the number of contacts is limited. The second term is greater than one due to the fact that every worker has to complete a wider range of tasks than before, and they lose the benefit of specialization. Because $n^*$ is the cost-minimizing communication choice of the firm, the second term dominates, and production costs increase with social distancing.

If the firm chooses to use telecommunication, the cost per contact will be $T>d^{\varepsilon}$ (otherwise the firm would have used telecommunication before). The proportional increase in production costs in this case is given by
\begin{equation}\label{telco}
\frac{c''(d)}{c(d)} 
	= T^\chi d^{\varepsilon \chi}
	>1.
\end{equation}
In both cases, the cost increase is highest for communication-intensive firms (large $\gamma$ and $\chi$) and those operating in a high-density area (high $d$ and hence high $n^*$).

\begin{figure}[!h]
\begin{tikzpicture}
\draw [<->] (0,3) node [above] {$c'/c$} --(0,0)--(4,0) node [right] {density};

\draw [dotted] (0,1) node [left] {$1$} --(4,1);

\draw (0,1)--(1,1);
\draw (1,1) .. controls (1.1,0.9) and (2.5,2) .. (4,2) node[right] {social distancing};

\draw (1,0.5) .. controls (1.1,0.4) and (2.5,2.5) .. (4,2.5) node[right] {telecommunication};
\end{tikzpicture}

\caption{{\bf Both social distancing and telecommunication hurt firms in dense areas more.}
See Eq \eqref{socdist} and Eq \eqref{telco} for the relative production costs under the two interventions.}
\label{fig2}
\end{figure}
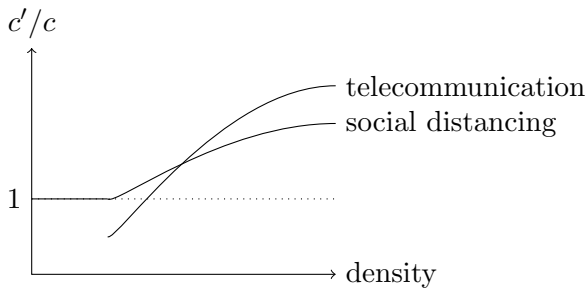

Fig \ref{fig2} displays the ratio of production costs under social distancing to the optimal production costs as a function of density. Firms in low-density areas are unaffected by social distancing since they do not have many contacts anyway. Those in intermediate-density areas would suffer less of a cost increase by switching to telecommunication. Firms in the highest-density areas will stick to face-to-face communication, which is still the more efficient form of communication despite restrictions. However, they will suffer the greatest cost increase. 

\section*{Data and methodology}
To estimate the potential disruptions from social distancing, we need a measure of the importance of worker interaction (corresponding to $\chi$ in the model) and its cost (captured by population density $d$). 

Let $\xi_o$ denote an indicator equal to one if occupation $o$ is interaction-intensive and zero otherwise. For industry $i$, $\chi_i = \sum_o s_{io}\xi_o$ measures the fraction of workers in affected occupations, with $s_{io}$ denoting the employment share of occupation $o$ in industry $i$.

We use the Occupational Information Network (O*NET) \cite{National_Center_for_ONET_Development2020-wj} to measure the characteristics of a given occupation, similarly to previous studies \cite{Firpo2011-hp,Autor2013-sh,Jin2020-tq}. The O*NET dataset contains detailed standardized descriptions on almost 1,000 occupations along eight dimensions. We focus on job characteristics that are related to recent social distancing measures, while prior work focused mainly on measuring offshorability of the given tasks \cite{Firpo2011-hp,Autor2013-sh}.

Social distancing interventions limit the interaction between people and regulate physical proximity between individuals. We thus focus on three related job characteristics based on work context and work activity described in O*NET. The first two indicators capture how communication-intensive the job is. Communication can be of two types: internal communication with co-workers (\emph{teamwork}) or external communication directly with customers (\emph{customer-facing}). The third indicator takes into consideration the possibility that workers may need to be in physical proximity of one another even if they do not communicate. We create an index that shows how important \emph{physical presence} is to perform a given job. Table \ref{table1} details the specific O*NET indexes that contribute to each of our three measures.
As social distancing measures only limit personal communication, for communication indexes, we require that the necessary face-to-face communication happens at least several times a week. 
In teamwork, face-to-face meetings can often be substituted by more structured communication, for which working from home is not as disruptive. To allow for this possibility, we only classify occupations as teamwork-intensive where both emails and letters and memos are less frequent forms of communication than face-to-face meetings. This excludes most managers and certain business services. Similarly, for physical presence, we require at least a certain degree of proximity to other workers which corresponds to working in a shared office.

\begin{table}[!ht]
\caption{
{\bf Definition of social distancing indexes.}}
\begin{tabular}{|l+l|l|}
\hline
{\bf Index} & {\bf Tasks} & {\bf Context} 
\\ \thickhline
\hline
Teamwork & Work With Work Group or Team & Face-to-face  \\
& Provide Consultation and Advice to Others & discussions  \\
& Coordinating the Work and Activities of Others & several times a week \\
& Guiding Directing and Motivating Subordinates & \& more often than \\
& Developing and Building Teams & emails, letters, memos\\
\hline
Customer & Deal With External Customers & Face-to-face\\
&  Performing for or Working Directly with the Public & discussions \\
& Assisting and Caring for Others & several \\
& Provide Consultation and Advice to Others & times a week\\
& Establishing and Maintaining Interpersonal Relationships &\\
\hline
Presence & Handling and Moving Objects & Density of \\
& Operating Vehicles, Mechanized Devices or Equipment & co-workers \\
& Repairing and Maintaining Electronic Equipment &  like shared\\
& Repairing and Maintaining Mechanical Equipment & office or more\\
& Inspecting Equipment, Structures, or Material &\\
\hline
\hline
\end{tabular}
\begin{flushleft} Each social distancing index (column 1) is created as an arithmetic average of the component indexes (column 2). To be classified an affected occupation, the average has to exceed 62.5 and the work context index has to exceed the threshold in column 3.
\end{flushleft}
\label{table1}
\end{table}

We aggregate the measures to 6-digit occupation codes (Standard Occupational Classification; 2010-SOC). We have information on the relevance of teamwork, customer contact and physical presence  for 809 occupations in SOC 2010 codes.

\begin{figure}[!h]
\plos{\includegraphics[width=0.7\linewidth]{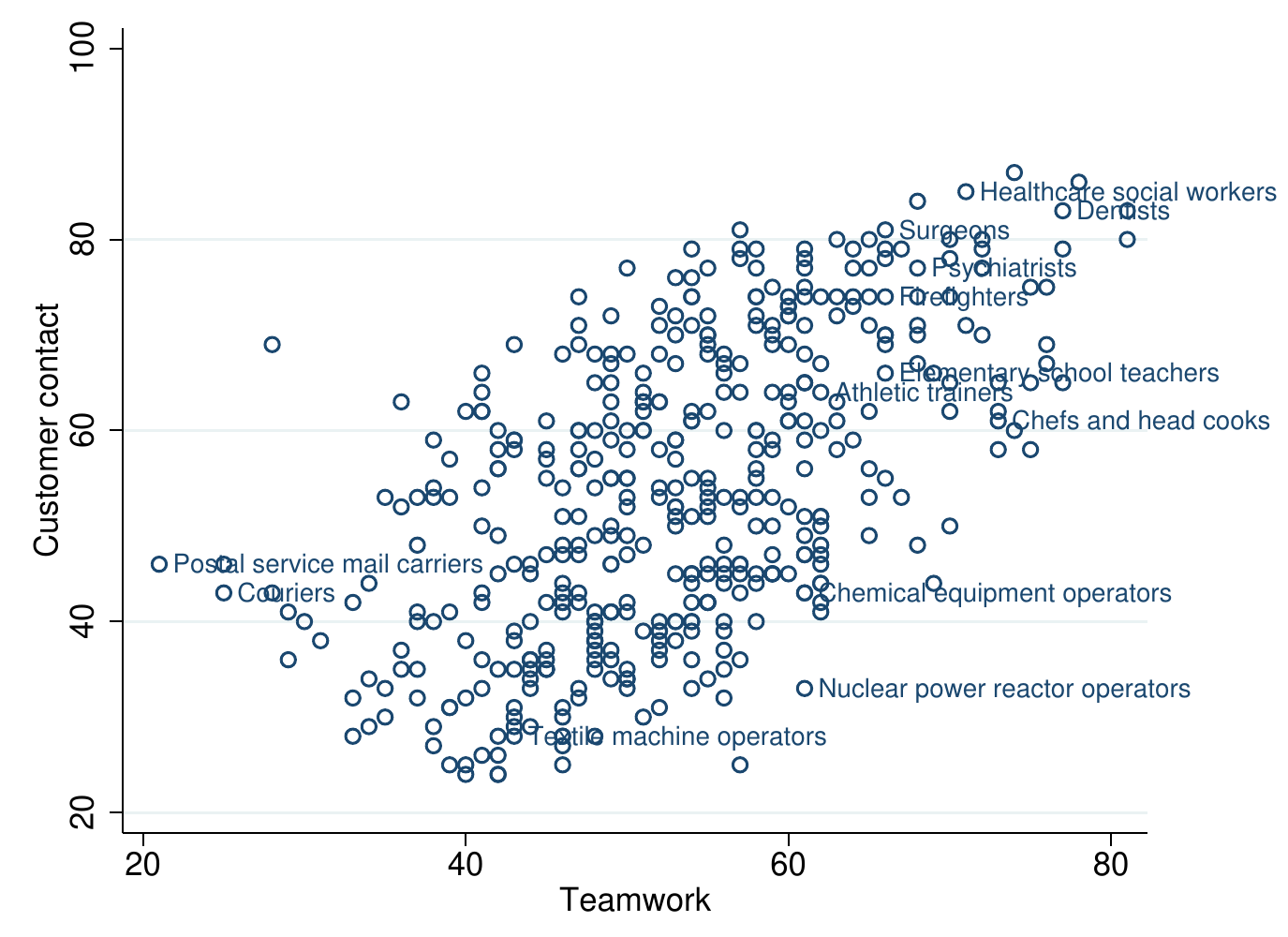}}

\caption{{\bf Teamwork and customer contact are highly correlated.}
Each circle represents an occupation. Teamwork and customer contact indexes are constructed as explained in main text.}
\label{fig3}
\end{figure}

Teamwork and customer contacts are highly correlated (Fig \ref{fig3}), but they are conceptually different. While all medical occupations require teamwork and customer contact, supervisors in general are working in teams but do not often communicate directly with customers. Machine operators and production workers in general are at the bottom of both of the distributions. As managers can substitute personal communication with emails, they are not considered in general as teamwork-intensive occupations according to our definition. Given the high correlation between the two types of communication, we often refer to \emph{communication-intensive} occupations that are either teamwork-intensive or customer-facing. 

As a next step, we calculate for each sector the share of workers whose job requires a high level of teamwork, customer contact and physical presence. 
We use the same sectoral breakdown as the Current Employment Statistics (CES) \cite{CES}. As all the indexes are an absolute value running from 0 to 100, we use 62.5 as a cutoff to define a job to be teamwork-intensive, customers contact-intensive or job that require physical presence from the worker.  The occupation structure of the industries are retrieved from the official industry-occupation matrix \cite{employment-matrix}, we use the employment statistics by occupation-industry for February 2020. 

Based on the share of relevant occupations in industry employment, the most teamwork-intensive sectors are, for example ``Hospitals,'' ``Accommodation'' and  ``Motion picture and sound recording industries.'' In contrast, teamwork is not important in sectors like ``Forestry and logging'' and ``Fishing, hunting and trapping.''  Customer contact is relevant in sectors like ``Hospitals''  and ``Retail'', while it is not relevant is sectors line ``Truck transportation,'' and ``Forestry and logging.'' Physical presence is relevant in sectors like  ``Truck transportation,'' ``Repair and maintenance,'' mining in general, while it is not relevant in finance and information technology sectors. 

``Hospitals'' score high on all three measures because communication in health care teams and with patients is important, and doctors and nurses work in close physical proximity to others. We nonetheless remove this sector from the analysis because of its inevitable direct role in combating the epidemic which is not captured well in a simple model of communication.

To measure the heterogeneity in the cost of communication, we measure population density in the neighborhood of businesses. The assumption in the model is that communication costs are lower in dense areas. This is consistent with the fact that communication-intensive sectors such as business services and advertising, together with central administrative offices of production firms tend to be concentrated in high-density areas \cite{Aarland2007-dh}.

The location of sectors comes from the County Business Patterns (CBP) data for 2017 \cite{CBP}. For a finer spatial resolution, we use the data tabulated by ZIP-Code Tabulation Areas. The CBP lists the number of establishments of a certain size for each ZIP-code and NAICS industry code. Because establishment sizes are given in bins (e.g., 1--4 employees), we take the midpoint of each bin as our estimated employment (e.g., 2.5 employees). In small industries and ZIP codes, the Census omits some size categories to protect the confidentiality of businesses. We impute employment in these plants from the national size distribution of plants in the same NAICS industry. Our estimated industry-level employment is a very good approximation to official employment statistics \cite{CES}. The correlation between our estimates based on CBP and the employment reported in CES is 0.98.

To understand the heterogeneity across regions, we average the share of workers in teamwork-intensive, customer-facing and physical presence occupations across industries active in each ZIP code. For each of the three occupation groups, let $\xi_{og}$ denote the indicator whether occupation $o$ belongs to group $g$. The industry share of workers in the occupation group is given by $\sum_o s_{io}\xi_{og}$. We compute the regional share of the occupation group as an employment-weighted average across industries, $\sum_i l_{ir}\sum_o s_{io}\xi_{og}/\sum_i l_{ir}$, where $l_{ir}$ is the estimated employment of industry $i$ in region (ZIP code) $r$. Because we only have industry but not occupation data by location, we have assumed that the occupation distribution within the sector, $s_{io}$, does not vary across locations. The fact that the CBP tabulates employment by establishments rather than firms makes this a good approximation. For example, a sporting good producer may also have an administrative office and a retail store in different locations, but these will be classified in their respective NAICS sectors rather than in sporting good manufacturing.

We use population density to measure how locations differ in the costs of communication. We also experimented with using employment densities instead of population densities. Results were very similar, as the two measures are highly correlated with the exception of very high employment-density urban centers where population is more sparse \cite{Heblich2017-lb}.

\subsection*{Counterfactual calculations}
To gauge the magnitude of the effect of social distancing, we calibrate the parameters of the model and compute the effect of a policy that limits the number of worker contacts. At the same time, we let the government introduce a proportional wage subsidy $\lambda$ to help offset the costs from lower interaction. With this subsidy, the cost of labor will be $(1-\lambda)$. We ask what level of $\lambda$ would compensate businesses for the communication disruption caused by social distancing. Using the cost change in Eq \ref{socdist}, we can express
\begin{equation}\label{subsidy}
    \lambda_{ir} = 1 -
        \frac{1-\chi_i}{1-\chi_i N/n_{ir}^*}
        \left(
        \frac{N}{n_{ir}^*}
        \right)^{\gamma_i}>0.
\end{equation}
The compensating wage subsidy increases in the importance of communication $\chi_i$ and the optimal number of contacts $n^*_{ir}$, and decreases in the number of allowed contacts $N$. The subscripts note that communication share is industry specific and the optimal number of contacts is both industry and region specific.

We calibrate the upper limit on personal contacts $N$ such that the overall number of contacts in the economy, when averaged across ZIP codes and industries, is reduced by half, $\sum_{i,r}l_{ir}\min\{N,n_{ir}^*\} = 0.5\sum_{i,r}l_{ir}n_{ir}^*$. Due to the inherent nonlinearity of the model, other interventions will have different effects.

To calibrate the importance of communication $\chi_i$, note that it is the cost share of communication, and can be correspondingly calibrated to the employment share of communication-intensive occupations in industry $i$. Here we take all occupations that are either teamwork intensive or customer facing.

Population density in region $r$ can be measured directly in the data, as explained above.
The only remaining parameter to calibrate is the elasticity of the communication externality $\varepsilon$. We rely on previous estimates of agglomeration effects that capture the elasticity of total factor productivity with respect to population density \cite{Ciccone1996-gu}. In our model, the elasticity of unit costs (which can be construed as the inverse of productivity) with respect to density is $-\varepsilon\chi$ (Eq \eqref{cost2}). We calibrate $\varepsilon=0.02$ so that, across all ZIP codes and industries, a regression of log model-implied productivity ($\varepsilon\chi_i\ln d_r$) on log population density $\ln d_r$ yields an elasticity of 0.04 \cite[page 60]{Ciccone1996-gu}.

Given these parameter values, we compute the compensating wage subsidy for each industry in each ZIP code using Eq \ref{subsidy}. We report employment-weighted averages of this across sectors and across locations.

\section*{Results}
Table \ref{table2} displays the top five and the bottom five industries by 2-digit NAICS industries as sorted by the percentage of workers in communication-intensive occupations, excluding hospitals and clinics. Across industries, retail trade, accommodation and food services, arts, entertainment, and recreation, other services and educational services have the highest share of communication-intensive workers, exceeding 35 percent. Transportation, production, construction and agricultural industries are less reliant on face-to-face communication. This heterogeneity across industries is important to understand the effect of social distancing measures.

\begin{table}[!ht]
\caption{
{\bf Retail, professional services, finance and restaurants are the most communication intensive.}}
\begin{tabular}{|l+c|c|c+c|}
\hline
 & \multicolumn{3}{c}{\bf Communication} &  
\\ 
{\bf Industry} & {\bf Teamw.} & {\bf Custom.} & {\bf Overall} & {\bf Presence} 
\\ \thickhline
Retail trade &13&67&68&5\\
Accommodation \& food services &8&50&51&1\\
Arts, Entertainment, and Recreation &12&40&42&2\\
Other Services (except Public Admin.) &12&38&41&12\\
Educational Services &15&35&37&1\\
...&&&&\\
Wholesale Trade &8&16&20&12\\
Transportation and Warehousing  &8&10&16&32\\
Manufacturing &7&6&11&10\\
Construction &15&5&18&28\\
Agri., forestry, fishing \& hunting &4&4&7&23\\
 \thickhline

\hline
\end{tabular}
\begin{flushleft} ``Teamw.'' and ``Custom.'' show the percentage of workers in teamwork-intensive and customer-facing occupations, respectively. ``Overall'' shows the percentage of workers in communication-intensive occupations that are either teamwork-intensive or customer-facing. It is less than the sum of the two indexes because some occupations rely on both types of communication. ``Presence'' shows the percentage of workers whose jobs require physical presence in close proximity to others.
\end{flushleft}
\label{table2}
\end{table}

\begin{figure}[!h]
\plos{\includegraphics[width=0.7\linewidth]{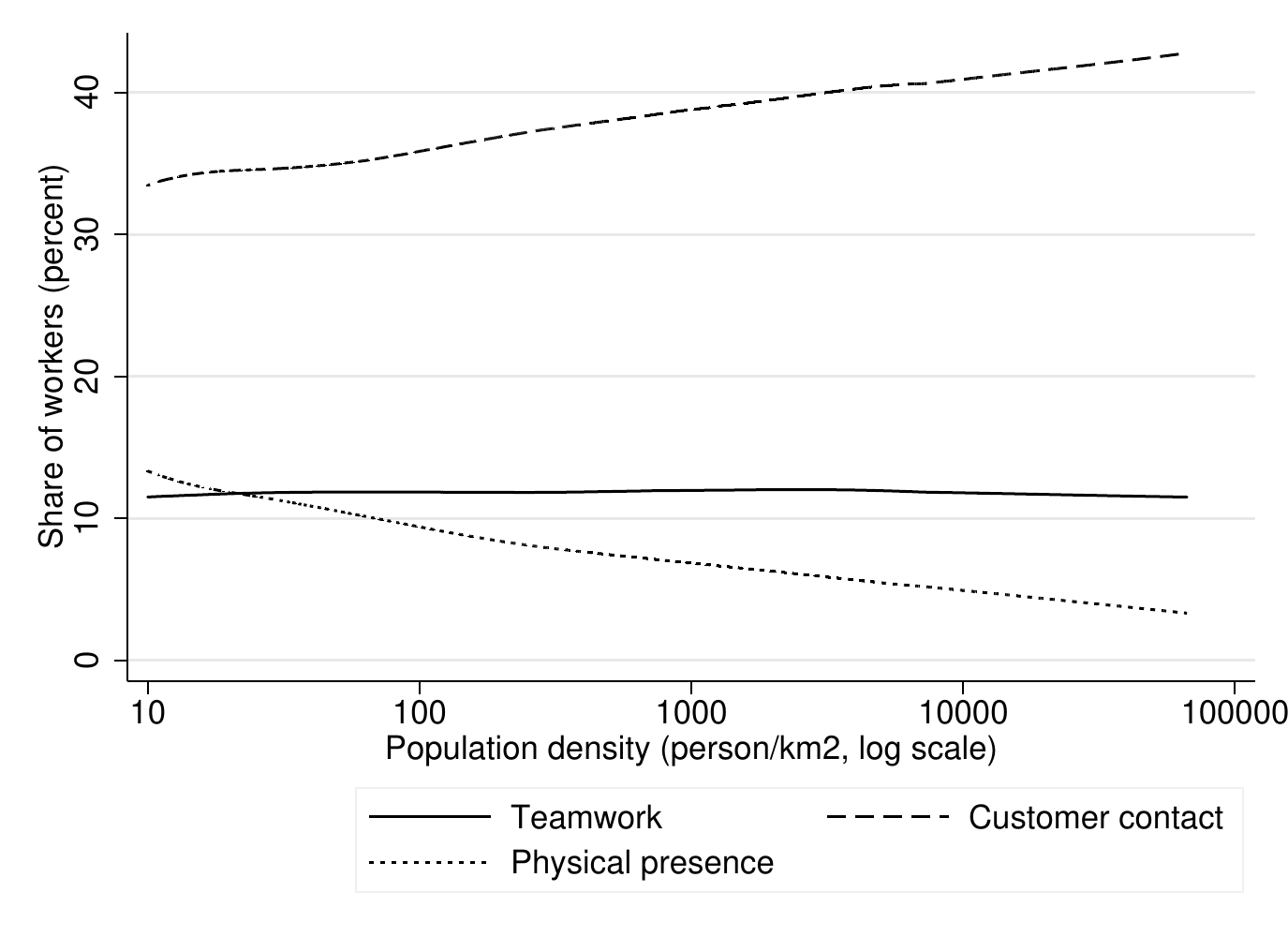}}

\caption{{\bf Urban areas employ more workers in customer-facing occupations, less in occupations requiring physical presence.}
Locally weighted polynomial regression of average share of teamwork-intensive occupations, customer-facing occupations and occupations requiring physical presence across sectors within the ZIP code (bandwidth $=0.5$).}
\label{fig4}
\end{figure}

Fig \ref{fig4} plots the share of workers in the three affected occupation groups across ZIP codes by population density. Customer-facing occupations are overrepresented in dense areas. In the highest population density ZIP codes, 43 percent of workers are employed in customer-facing occupations. Teamwork-intensive occupations are broadly distributed in space.

In the calibrated model, a social distancing policy that puts a cap on interactions per worker such that the total number of interactions drops by half nationwide is compensated a 12.2 percent wage subsidy. The distribution of the compensating wage subsidy is, however, unequal over space and across industries. 
New York City, with a population density about 20 times the average U.S. city, would require a 13.3 percent wage subsidy. 
By contrast, the compensating wage subsidy in agriculture, transportation and manufacturing would be less than 6 percent (Table \ref{table3}).

\begin{table}[!ht]
\caption{
{\bf The five most affected sectors require more than 20 percent wage subsidy.}}
\begin{tabular}{|l+c|c|}
\hline
{\bf Industry} & {\bf Wage subsidy} & {\bf Employment} 
\\ \thickhline
Retail Trade	&22.1	&15,659\\
Accommodation and Food Services	&17.7	&14,379\\
Arts, Entertainment, and Recreation	&15.1	&2,494\\
Other Services (except Public Admin.)	&14.5	&5,939\\
Educational Services	&13.8	&3,838\\
...&&\\
Wholesale Trade	&7.7	&5,936\\
Construction	&6.8	&7,646\\
Transportation and Warehousing	&5.9	&5,523\\
Manufacturing	&4.5	&12,861\\
Agriculture, Forestry, Fishing and Hunting	&2.6	&55\\
\hline
{\bf Average} & {\bf 12.2} & {\bf 116,496}\\
 \thickhline

\hline
\end{tabular}
\begin{flushleft} ``Wage subsidy'' displays the percentage decrease in labor costs necessary to compensate businesses when worker contacts are reduced by half. ``Employment'' is the February 2020 employment of the sector in thousands \cite{CES}. The last row shows the employment-weighted average wage subsidy. Table excludes hospitals, clinics, and government establishments which are not present in CBP.
\end{flushleft}
\label{table3}
\end{table}

\section*{Discussion and conclusions}
The main cost of social distancing in our model is insufficient division of labor. This mechanism is motivated by \cite{Smith1778-qq} and captures the same trade-off as \cite{Becker1992-ac}. Our contribution is specifying the cost function in such a way that can be easily mapped to the data.

More broadly, our argument is that frequent interaction increases productivity irrespective whether it is happening between workers, between workers and customers, or between workers and machines. In the main part of the empirical analysis, we focused only on the first two types of interactions, while we were silent about the third. But social distancing measures also affect sectors where workers need to be in physical proximity of one another even if they do not communicate, for example, to operate machinery or access key resources. This is relevant in sectors like ``Mining, Quarrying, and Oil and Gas Extraction'' and ``Transportation'' while it is not relevant in sectors like ``Finance and Insurance'' and ``Professional, Scientific, and Technical Services.'' As can be seen from Fig \ref{fig4}, occupations requiring physical presence have the highest share in less dense areas where production and mining plants are located. (Farms are not included in the CBP.) The share of these occupations in the most dense areas is only 3 percent.


To a greater or lesser extent, all sectors will be affected by social distancing. Some sectors are hit by the intervention due to restricted face-to-face communication, others are hit due to restricting physical proximity of people. Some sectors are less affected across all dimensions. Examples include ``Fishing, hunting and trapping,'' ``Printing and related support activities,'' and manufacturing in general. 

We see four avenues for further research. The first concerns the demand side of the economy, which we have mostly neglected by focusing on the production function. The employment response to a production disruption greatly depends on the elasticity of demand. We hypothesize that sectors like schooling and health care have inelastic demand and will continue to employ many workers despite significant disruptions. However, personal services, small grocery stores and restaurants may face more elastic demand and respond to large production cost increases by laying off a significant fraction of their work force.

The second direction concerns the interaction between sectors and regions. Whenever productivity in any business drops, this shock can propagate to its buyers and suppliers. The aggregate consequences of the epidemic will hence be modulated by input-output linkages between sectors, regions and countries \cite{Caliendo2014-mr,Caselli2020-nf,Baldwin2020-wb}.

The third and forth directions concern the long-run response of businesses as they try to become more resilient to such shocks in the future. Whether the share of telecommunication remains large in the long run depends crucially on how easily it substitutes for face-to-face interaction. In our model, the two are perfect substitutes with the only distinction that the efficiency of face-to-face meetings improves with population density, whereas telecommunication does not depend on agglomeration \cite{Rossi-Hansberg2007-tm,Ioannides2008-bs,Tian2019-wq}. Previous work has found face-to-face communication to be more effective in high-intensity communication which is particularly helpful to overcome incentive problems in joint production \cite{Gaspar1998-gy,Storper2004-mg}. Data on internet flows suggests that telecommunication is not a good substitute for face-to-face meetings \cite{Cuberes2013-js}. None of these papers discuss disruptions from social distancing measures. Further study of the modes of communication in the O*NET occupation survey can shed light on whether telecommunication can act as a low-cost substitute for face-to-face meetings.

Fourth, businesses may change their location in response to perceived threats and disruptions. As we discussed, epidemics have a disproportionate effect on cities. So it is conceivable that in a post-pandemic spatial equilibrium (not modeled here, but see \cite{Tian2019-wq}), the agglomeration premium falls and firms find it less attractive to locate in cities. A poignant point of comparison is the increased threat of terrorism in major cities following devastating attacks on New York, Washington, London, Paris, Madrid, Moscow and Mumbai. The general conclusion about terror threat is that cities have remained resilient and a robust attractor of businesses \cite{Glaeser2002-mw,Harrigan2002-ik}. We speculate that epidemics and social distancing can be more detrimental to cities than terror threats, because they tear apart the very fabric of urban life. However, we have limited data to make further predictions.

\section*{Supporting information}

\paragraph*{S1 Full table of sectors.}
\label{S1_File}
{\bf Social distancing exposure by sector.} Available at \url{https://github.com/ceumicrodata/social-distancing/blob/master/data/derived/industry-index.csv}.

\paragraph*{S2 Full table of ZIP-codes.}
\label{S2_File}
{\bf Social distancing exposure by location.} Available at \url{https://github.com/ceumicrodata/social-distancing/blob/master/data/derived/location-index.csv}.

\paragraph*{S3 Data repository.}
\label{S3_URL}
{\bf Replication code and data.} Replication code and data are available at \url{https://github.com/ceumicrodata/social-distancing}.

\section*{Acknowledgments}
We thank Gábor Békés, Péter Harasztosi, Péter Karádi, Balázs Lengyel, Dávid Krisztián Nagy, and Andrea Weber for comments.

\nolinenumbers




\end{document}